# Coulomb blockade of field emission from nanoscale conductors

O. E. Raichev*

*Institute of Semiconductor Physics, National Academy of Sciences of Ukraine, Prospekt Nauki 45, 03028, Kiev, Ukraine*

Theoretical description of the field emission of electrons from nanoscale objects weakly coupled to the cathode is presented. It is shown that the field-emission current increases in a steplike fashion due to single-electron charging which leads to abrupt changes of the effective electric field responsible for the field emission. A detailed consideration of the current-voltage characteristics is carried out for a nanocluster modeled by a metallic spherical particle in the close vicinity of the cathode and for a cylindrical silicon nanowire grown on the cathode surface.

PACS number(s): 79.70.+q, 73.23.Hk, 73.40.Gk

## I. INTRODUCTION

The discrete nature of electric charge reveals itself in the transport of electrons through small conductors (nanoparticles or other nanoscale objects) weakly coupled to the source and drain electrodes (current-carrying leads) owing to the Coulomb blockade effect. Numerous manifestations of the charge quantization in transport properties, the most familiar of them are the Coulomb blockade oscillations of the electric current as a function of the gate voltage and the Coulomb staircase in the current-voltage characteristics, have attracted considerable attention in the past years.[1] Since the fundamentals of the transport theory in the Coulomb blockade regime have been established,[2–4] the Coulomb blockade-based physics has been applied to various issues of electron transport in mesoscopic systems, and the field of its applications expands in line with the advances in nanotechnology.

Usually, the influence of the Coulomb blockade on the current in two-terminal devices is considered under assumption that the coupling between the nanoscale object and the leads is not sensitive to the number of electrons $N$ determining the object charge $eN$. This corresponds to the introduction of ohmic (or nearly ohmic) effective resistances describing this coupling. Though this assumption often works well, it can be violated, for example, in nanomechanical systems,[5–7] where charging of the object gives rise to its displacement towards one of the leads thereby changing its tunnel coupling to both leads. In this paper we study a situation when the sensitivity of the tunnel coupling to the number of electrons does not require a mechanical displacement and is determined by the nature of tunneling. This implies a device layout and conditions similar to those used in the recent experiments on field emission of electrons from metallic nanoclusters,[8–10] silicon nanowires[11–15] and nanocones,[16,17] and carbon nanotubes (see, for example, Refs. 11 and 18–26), when small (nanoscale) objects are formed on the source electrode (cathode), the latter is then negatively biased with respect to the drain electrode (anode) in vacuum. The current between the electrodes flows owing to the field emission of electrons from nanoscale objects, because the electric field $F$ at the tips of the objects is higher than in the other places of the device. The field-emission current is described by the Fowler-Nordheim formula[27]

$$I = ASF^2 \exp\left(-\frac{\mathcal{F}}{F}\right), \quad \mathcal{F} = \frac{4\sqrt{2m}}{3|e|\hbar} W^{3/2}, \qquad (1)$$

where $m$ is the free electron mass, $W$ is the work function of the emitting material, $S$ is the effective emitting area, and $A$ is a constant expressed through the work function and Fermi energy $\varepsilon_F$ of the emitting material

$$A \simeq \frac{|e|^3 \sqrt{\varepsilon_F/W}}{4\pi^2 \hbar (\varepsilon_F + W)}. \qquad (2)$$

The effective field $F$, which describes the tunnel coupling between the nanoscale object and the anode, depends on the object charge, which is induced by the applied voltage $V = V_1 - V_2$, where $V_1$ and $V_2$ are the cathode and anode potentials, respectively. Under conditions of Coulomb blockade, i.e., when the electric connection between the cathode and the object is weak and the charging energy of the object considerably exceeds the temperature $T$, the *continuous* variation of the voltage $V$ leads to *discrete* changes of the object charge in units of $e$, and, consequently, to corresponding discrete changes of the field $F$. Therefore, one may introduce the field $F_N$, which is a function of the discrete number $N$ and continuous variable $V$. Next, if the current in the device is limited by the field emission, the single-electron tunneling processes become important. This means that, at a fixed voltage $V$, the object stays mostly in the states with $N$ and $N-1$ electrons, the number $N$ is determined by the voltage. In the $N$-electron state, no electrons can come to the object from the cathode until an electron leaves the object by tunneling through the barrier, see Fig. 1(a). Then the object appears in the $N-1$-electron state and returns to the $N$-electron state before the next Fowler-Nordheim tunneling event takes place. The field-emission current in these conditions is given by Eq. (1) with $F = F_N$ and can be denoted as $I_N$. If the bias $eV$ increases, the state with $N+1$ electrons becomes more favorable, and the current changes in a steplike fashion from $I_N$ to $I_{N+1}$. This leads to staircaselike current-voltage characteristics, which may look similar to the usual Coulomb staircases.[28–30] However, since the sensitivity of the tunneling to the number of electrons is involved, the staircaselike current-voltage characteristics can exist under rather peculiar conditions, when the source-drain bias is orders of magnitude larger than the charging energy.

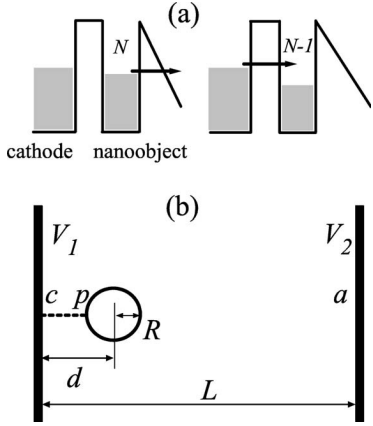

FIG. 1. (a) The mechanism of single-electron tunneling in the Fowler-Nordheim regime. (b) Schematic representation of the idealized emitter.

The rest of the paper is devoted to quantitative studies based on the physical idea outlined above. In Sec. II we give the basic equations and calculate the current in the simplest case of an idealized emitter shown in Fig. 1(b). In Sec. III we calculate the current from a nanocluster modeled by a spherical particle on the metallic cathode surface and from a semiconductor wire (nanowhisker) grown perpendicular to the cathode surface. The discussion and concluding remarks are given in the last section.

## II. GENERAL CONSIDERATION

We consider the case of classical (or metallic) Coulomb blockade, when the electron energy level separation in the nanoscale object can be neglected in comparison to both temperature and charging energy. Since the object is assumed to be weakly coupled to the cathode, we study the sequential tunneling process and not the coherent one. It is convenient to investigate the electron transport by applying the kinetic equation[2] (Master equation) for the distribution function $P_N$ describing the probability for the object to be in the state with $N$ electrons. Assuming that the electric connection between the cathode and the object is characterized by the conductance $G$, this equation is written as

$$\frac{\partial P_N}{\partial t} = Q_{N+1} - Q_N, \quad (3)$$

where

$$Q_N = \frac{G}{e^2} \frac{\Delta E_N}{1 - \exp(-\Delta E_N/T)} [P_N - P_{N-1} \exp(-\Delta E_N/T)] + P_N I_N/|e|. \quad (4)$$

Here $\Delta E_N = (e^2/C)[N - 1/2 - C_2 V/e]$ is the difference in Coulomb energies for the objects with $N$ and $N-1$ electrons, $C$ is the total capacitance, and $C_2$ is the capacitance of the object with respect to the anode (the capacitance with respect to the cathode is given by $C_1 = C - C_2$). The first term in expression (4) has the usual form[2] and corresponds to the current between the object and the cathode. It is written as a difference of the contributions describing the departure of an electron from the object in the $N$-electron state and arrival of an electron at the object in the $N-1$-electron state. The second term corresponds to the field-emission current from the object in the $N$-electron state. Since no electrons come to the object from the anode, this term does not contain a contribution describing arrival of electrons. In the stationary case, Eq. (3) is reduced to the form $Q_N = \text{const}$, where the constant can be chosen equal to zero. After determining $P_N$ from the equation $Q_N = 0$ with the use of the normalization condition $\Sigma_N P_N = 1$, the total current is given by

$$J = \sum_N P_N I_N. \quad (5)$$

Under the condition $GT \gg |e| I_N$, which means that the object is in thermal equilibrium with the cathode, the stationary solution of Eq. (3) is written as $P_N = Z^{-1} \exp(-E_N/T)$, where $E_N = (e^2/2C)[N - C_2 V/e]^2$ is the Coulomb energy, and $Z = \Sigma_N \exp(-E_N/T)$ is the partition function. The current in this case is determined by the expression

$$J = Z^{-1} \sum_N I_N \exp(-E_N/T). \quad (6)$$

Let us apply the solution (6) to the idealized model of emitter, Fig. 1(b), when the emission takes place from a spherical nanoparticle of radius $R$, placed at a distance $d$ from the cathode. The distance between the cathode and anode is $L$. The connection $c$-$p$ denotes a low-transparent contact (for example, tunnel barrier) between the particle and the cathode, which does not contribute to the field-emission properties and electrostatics of the device. Assuming $d \gg R$, we have $C = R$, $C_2 = Rd/L$, and neglect the charge polarization of the particle because this polarization is small in comparison to the total charge $eN$ induced by the applied voltage. The number of electrons is estimated as $N \simeq C_2 V/e = Rd F_0/|e|$, where $F_0 = -V/L$ is the applied electric field. The effective field for the nanoparticle with $N$ electrons is $F_N = |e|N/R^2$, and the partial currents $I_N$ in these conditions are given by

$$I_N = AS(eN/R^2)^2 \exp(-\mathcal{F}R^2/|e|N), \quad (7)$$

where the emitting area $S$, in the idealized model considered here, can be approximated by the total surface area of the nanoparticle, $S = 4\pi R^2$. In Fig. 2 we plot the current-voltage characteristics of the idealized emitter, calculated according to Eqs. (6) and (7), where $A$ is given by Eq. (2) with $W = 5.1$ eV and $\varepsilon_F = 5.5$ eV (taken for Au), and the geometrical parameters are chosen as $R = 5$ nm and $d = 0.5$ $\mu$m. The characteristics look like staircases with flat regions (plateaus) between the steps, which are visible even at room temperature. It is possible to estimate the relative heights of the steps by calculating the ratio of the currents $I_N$ and $I_{N-1}$ emitted from the nanoparticle with $N$ and $N-1$ electrons

$$\frac{I_N}{I_{N-1}} \simeq \exp\left[\frac{\mathcal{F}R^2}{|e|N(N-1)}\right]. \quad (8)$$

In spite of the fact that the charged nanoparticle typically contains a large number of electrons, $N \sim 100$, one can al-ways find a regime when the ratio $I_N/I_{N-1}$ is not small in comparison to unity. This necessarily implies a weak Fowler-Nordheim tunneling, when $\mathcal{F}/F = \mathcal{F}R^2/|e|N \gg 1$.

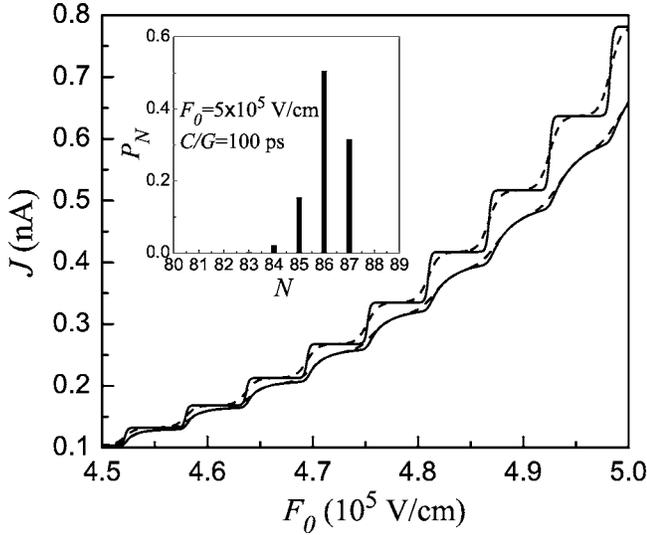

FIG. 2. Current from the idealized emitter as a function of the applied field $F_0 = -V/L$ for the case of small $C/G$ (nanoparticle in thermal equilibrium with the cathode, upper curves) and for the case of $C/G = 100$ ps (lower curves), at the temperatures $T=77$ K (solid) and $T=293$ K (dashed). The inset shows the distribution function $P_N$ at $F_0 = 5 \times 10^5$ V/cm for the second case.

In the calculations described above, the applicability of the Fowler-Nordheim formula requires $R \gg W/|e|F$, which is rewritten as $R \ll e^2 N/W$, or, according to $N \simeq R d F_0/|e|$, as $|e|F_0 \gg W/d$, independent of the nanoparticle radius. This condition is satisfied at high enough applied voltages. If $|e|F_0 = eV/L \lesssim W/d$, the approximation of a triangular potential barrier is not quite good, and one should consider the tunneling through the barrier described by the potential energy $W - e^2 N(1/R - 1/r)$ at $r \geq R$, where $r$ is the distance from the center of the spherical nanoparticle; the tunneling through the potential barrier of this form is described in Ref. 31. Even under the condition $|e|F_0 \gg W/d$, which is satisfied in the calculations shown in Fig. 2, the relative change of the current per one step, $I_N/I_{N-1} - 1$, appears to be significant, because the exponent $\mathcal{F}R^2/|e|N(N-1)$ in Eq. (8) is estimated as $c(W/|e|F_0 d)^2$, where the dimensionless constant $c = 4/3\sqrt{2me^4/\hbar^2 W}$ is noticeably larger than unity.

If the current is high enough, the field emission cannot remain the bottleneck for the electron transfer from the cathode to the anode, and a finite resistance $G^{-1}$ becomes essential. The nanoparticle in these conditions is no longer in equilibrium with the cathode. This means that the distribution $P_N$ is established kinetically, and several states with different charges coexist at a fixed voltage (see the inset in Fig. 2). As a consequence, the Coulomb blockade features are washed out. This case requires a numerical solution of the equation $Q_N = 0$. The corresponding current-voltage characteristics of the idealized emitter calculated by using the RC time $C/G = 100$ ps are also shown in Fig. 2. The degradation of the current steps appears to be stronger with increasing voltage, because the current increases and the nanoparticle-cathode link becomes more important. The shape of the steps in this case resembles the usual Coulomb staircase.

## III. MORE COMPLEX EXAMPLES

After demonstrating the possibility of the Coulomb-blockade staircase of the field emission on a model example, it is worth to consider more complex cases. Indeed, the model example discussed above has certain disadvantages. First of all, it is hardly possible to connect a particle placed far from the cathode surface by a link [c-p in Fig. 1(b)] which does not contribute to the electrostatic properties of the device. Second, the model of uniform charging is insufficient: the charge polarization of the nanoscale object appears to be important and should be always taken into account, see below in this section. Therefore, the model shown in Fig. 1(b) is suitable only for the purposes of illustration of the basic physics described by Eqs. (3)–(6). To have a closer approach to reality, we point out that the nanoscale objects investigated in the above-cited experiments on field emission can be roughly divided into two classes: the objects whose dimensions in all directions are comparable (nanoclusters or nanoparticles), and the objects whose length in the direction of the applied field is much larger than their transverse size (nanowires or nanowhiskers). The following consideration is carried out for the cases of nanoclusters and nanowires of the simplest geometries, when the electric fields $F_N$ and the capacitances $C$ and $C_2$ can be determined consistently by solving corresponding electrostatic problems. The current is calculated according to Eq. (6), under the assumption that the objects are in equilibrium with the cathode.

### A. Field emission from nanoclusters

Below we consider the field emission from a nanocluster modeled by a spherical metallic particle of radius $R$ deposited on the flat cathode surface. To provide a finite capacitance $C$, one should assume a finite separation $d-R$ between the particle and the metallic cathode plate (for instance, one can imagine that the particle resides on an oxidized surface), see the inset to Fig. 3. Besides, this assumption provides electrical isolation of the particle from the cathode, which is a necessary condition for the Coulomb blockade. The electrostatics of the plane-sphere system is known, and the field and charge distributions in this case can be found in the form of rapidly converging infinite series arising from the potentials of image point charges and point dipoles.[32] Such a consideration allows one to present the distribution of the electrostatic potential energy near the particle in the approximate form

$$U(r,\theta) \simeq W + e\{\beta F_0 R[1 + \gamma(\cos\theta - 1)]$$
$$- \lambda[eN - C_2 V]/C\}\frac{r-R}{r}, \quad (9)$$

where $r$ and $\theta$ are the radial and azimuthal coordinates of the spherical coordinate system with the origin at the center of the particle, and $\beta$, $\gamma$, and $\lambda$ are the dimensionless constants of the order of unity, which are to be determined from numerical calculations. Such calculations also give us the capacitances $C$ and $C_2$.[33] Note that if the charge quantization is neglected (so that $N=C_2V/e$ when the particle is in equilibrium with the cathode), $\beta$ is identified with the field enhancement factor conventionally used in the physics of field emission. The expression (9) provides an excellent description of the electrostatic potential at $r-R<R/2$ and at small $\theta$. It allows one to take into account deviations of the potential energy from the linear form $W-|e|F(r-R)$ and, therefore, to find corrections to the Fowler-Nordheim tunneling exponent. Neglecting such corrections in the prefactor, we obtain the following expression for the partial currents

$$I_N = ASF_N^2 \exp\left[-\frac{\mathcal{F}}{F_N}\Phi\left(\frac{|e|F_N R}{W}\right)\right],$$

$$\Phi(x) = \frac{3}{2}\left[\frac{x^2}{\sqrt{x-1}}\left(\frac{\pi}{2} - \arctan\sqrt{x-1}\right) - x\right], \quad (10)$$

where $A$ is given by Eq. (2), the dimensionless function $\Phi(x)$ describes the corrections to the tunneling exponent, and the effective emitting area $S=2\pi R^2(F_N^2/\gamma\mathcal{F}\beta F_0) \simeq 2\pi R^2(F_N/\gamma\mathcal{F})$ is reduced due to the angular dependence of the radial field described by Eq. (9). The field $F_N$ is given by

$$F_N = \beta F_0 + \lambda\frac{|e|N - \tilde{C}_2 F_0}{CR}, \quad (11)$$

where the quantity $\tilde{C}_2 = C_2 L$ does not depend on the distance $L$ between the cathode and anode. Note that, since we always assume that $L$ is much larger than any dimension of the nanoscale object, the capacitance $C_2$ is always proportional to $1/L$, and it is more convenient to replace $C_2|V|$ by $\tilde{C}_2 F_0$. This substitution also allows us to represent the Coulomb energy standing in Eq. (6) as

$$E_N = \frac{e^2}{2C}[N - \tilde{C}_2 F_0/|e|]^2. \quad (12)$$

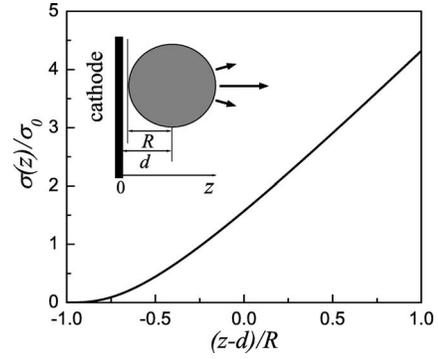

FIG. 3. Charge density per unit length in $z$ direction for a spherical metallic nanocluster placed at the distance $0.1\,R$ from the metallic cathode. Here $\sigma_0 = F_0 R/2$. The inset shows the geometry of the problem and the directions of the field emission (arrows).

Further calculations are done for the separation $d-R=0.1R$, when $C=2.16R$, $\tilde{C}_2=1.74R^2$, $\beta=4.32$, $\lambda=1.22$, and $\gamma=0.66$. Figure 3 shows the distribution of negative charges on the surface of the spherical particle staying in equilibrium with the cathode for this case ($|e|N=\tilde{C}_2 F_0$ is assumed). The distribution of the radial field $F(z)$ at the surface of the particle is given by the same dependence, $F(z)/F_0=\sigma(z)/\sigma_0$.

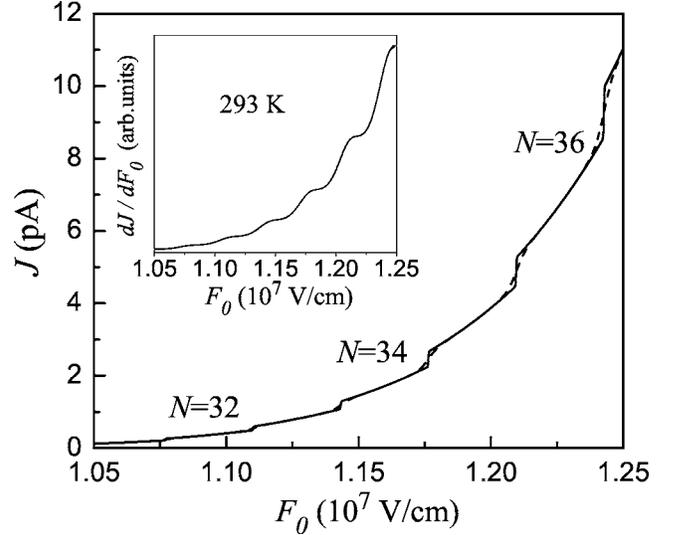

FIG. 4. Current from the spherical nanocluster of radius $R=5$ nm as a function of the applied field $F_0=-V/L$ at $T=4.2$ K (solid) and 77 K (dashed). The inset shows the derivative of the current at $T=293$ K.

The field-emission current from the nanocluster described above has been calculated according to Eqs. (6) and (10)–(12) at $R=5$ nm. The results of the calculations shown in Fig. 4 demonstrate the staircaselike behavior caused by the Coulomb blockade. However, in contrast to the staircases shown in Fig. 2, the current continues to increase between the steps. This occurs because of electrostatic polarization of the nanoparticle. According to Eq. (11), when the particle charge is constant, the increase in the applied field $F_0$ leads to an increase in the effective field $F_N$ because the factor $\beta - \lambda\tilde{C}_2/CR$ is positive. For the chosen particle radius, the steps of the current are clearly visible at liquid nitrogen temperature but poorly visible at room temperature. Nevertheless, the Coulomb blockade features at room temperature become quite distinct in the plots of the derivative of the current, as shown in the inset to Fig. 4.

### B. Field emission from nanowires

Let us consider the field emission from a small semiconductor wire modelled by a cylinder of radius $R$ and length $d$, which ends with a hemispherical tip of the same radius, see the inset to Fig. 5. The cathode substrate upon which the wire is grown is assumed to be a metal (or a heavily doped semiconductor) so that one can use the method of image charges instead of solving the electrostatic problem in the whole space. The electric isolation of the wire from the cathode in this case takes place in a natural way, because a Schottky barrier is formed between the wire and the metallic cathode (in the case of semiconducting cathode there can be a heterobarrier or an interband $p$-$n$ barrier). In other words, the wire region adjacent to the cathode becomes depleted of electrons and positively charged because of the presence of donors (we assume that the wire is uniformly doped with bulk donor density $n_D$). When a bias $eV$ is applied between the cathode and anode, the wire acquires a considerable negative charge because of tunneling or thermionic emission of electrons from the cathode through the barrier. When the field emission from the wire of nanoscale radius becomes essential, the density of induced negative charges per unit length of the wire appears to be much larger than the equilibrium charge density $\sigma_D = \pi R^2 |e| n_D$ even if $n_D$ is of the order of $10^{18}$ cm$^{-3}$. For this reason, one can use the "metallic" approximation assuming that the charges in the wire are placed mostly on its surface. This means that the electron density distribution $n(\rho,z)$, which depends on the radial coordinate $\rho$ of the cylindrical coordinate system connected with the wire, is given by $n(\rho,z)=(2\pi\rho|e|)^{-1}\delta(\rho-R)\sigma(z)+n_D$ for $z \leq d$ and $n(\rho,z)=(2\pi\rho|e|)^{-1}\delta[\rho-\sqrt{R^2-(z-d)^2}]\sigma(z)+n_D$ for $d \leq z \leq d+R$, where $\sigma(z)$ is the density of negative charges on the surface per unit length. Since this approximation is based on the assumption that the screening length is small in comparison with the wire radius, it works better for wider wires. For silicon wires, whose field-emission properties are currently the subject of investigations,[11–15] the metallic approximation remains suitable even for the radius of several nanometers, because, owing to the large effective masses and six-valley degeneracy, the density of electron states in $n$-Si appears to be high enough to provide the Thomas-Fermi screening length less than one nanometer for Fermi energies $\varepsilon_F > 0.01$ eV. The metallic approximation, of course, fails to describe the region of the wire in the close vicinity of the cathode, where the depletion occurs. Nevertheless, since this region is a small part of the whole wire, see the charge distribution in Fig. 5, its presence cannot considerably modify the parameters calculated as described below.

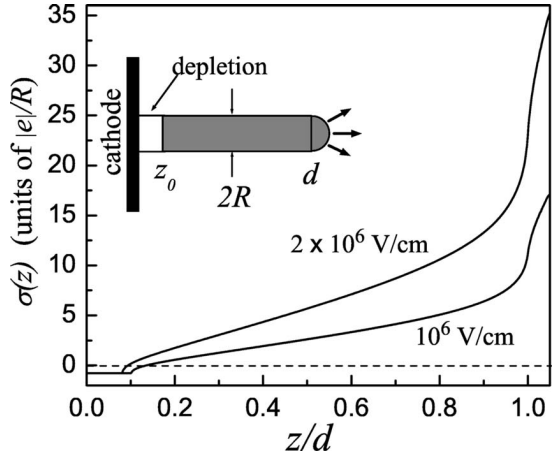

FIG. 5. Charge density per unit length for the cylindrical wire whose geometry is shown in the inset (see parameters in the text).

According to the discussion given here, we search for the charge distribution $\sigma(z)$ satisfying the integral equation

$$U(z) = U_0 - |e|F_0 z + \int_0^{d+R} dz' K(z,z')\sigma(z'), \quad (13)$$

where $U(z)$ is the potential energy counted from the Fermi level in the cathode material, $U_0$ is the barrier height, and $K(z,z')$ is the potential of interaction between the electrons in the points $z$ and $z'$ of the wire surface in the presence of the cathode plate, see the Appendix. Equation (13) is accompanied with additional requirements: $U(z)=0$ at $z>z_0$ and $\sigma(z)=-\sigma_D$ at $z<z_0$, where $z_0$ is the depletion edge coordinate, which is to be found self-consistently. The first of these requirements corresponds to a full screening of the bare potential $U_0-|e|F_0 z$ by the induced charges of the wire, while the second one models the presence of the positive charges in the depletion region $z<z_0$. Once the distribution $\sigma(z)$ is found, the total charge of the wire, $-\int_0^{d+R} dz\sigma(z)$, as well as the distribution of electric field around the wire, can be calculated. To find the capacitance $C$ and describe modification of the effective field under single-electron charging, one may calculate the variation of the total charge and the field at the tip (at $z=d+R$) with respect to a small variation of $U_0$. Equation (13) is solved numerically by using the method of iterations. The dependence of the effective field $F_N$ on $F_0$ and $N$ can be represented in the form similar to Eq. (11)

$$F_N = \beta(F_0)F_0 + \lambda \frac{|e|(N+B) - \tilde{C}_2 F_0}{C(F_0)R}, \quad (14)$$

while the Coulomb energy is written as

$$E_N = \frac{e^2}{2C(F_0)}[N + B - \tilde{C}_2 F_0/|e|]^2. \quad (15)$$

These equations take into account a finite (though weak) dependence of the capacitance $C$ and field enhancement factor $\beta$ on the applied field $F_0$. The dependence of the parameters $\tilde{C}_2$ and $\lambda$ on $F_0$ appears to be much weaker and can be neglected. The positive dimensionless constant $B$ reflects the fact that the average number of induced charges is smaller than $\tilde{C}_2 F_0/|e|$. These features appear because the system under consideration is not entirely metallic and contains a depletion region whose length changes with $F_0$.

The numerical calculations leading to the results presented below are done for $U_0=0.7$ eV, which approximately corresponds to the Schottky barrier height for $n$-Si in contact with Al.[34] The chosen donor density is $n_D=2\times10^{18}$ cm$^{-3}$. The parameters standing in Eqs. (14) and (15), however, are not sensitive to $n_D$, except for the capacitance $C$, which changes within 10% when $n_D$ varies from $10^{18}$ cm$^{-3}$ to $2\times10^{18}$ cm$^{-3}$. Figure 5 shows the charge density distribution for the wire of radius $R=5$ nm and length $d=0.1$ $\mu$m at $F_0=10^6$ and $2\times10^6$ V/cm. The charge density shows a nearly linear growth through the main part of the wire and a sharp enhancement at the hemispherical tip from which the field emission occurs. The dependence of the field enhancement factor and capacitance on the applied electric field is shown in Fig. 6, and the other calculated parameters are $\tilde{C}_2=2.44\ dR$, $\lambda=0.414$, and $B=12.14$.

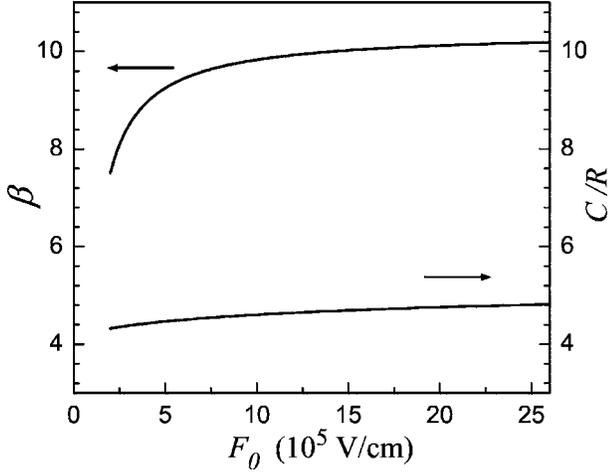

FIG. 6. Field dependence of the enhancement factor and capacitance for the cylindrical wire with $R=5$ nm and $d=0.1$ $\mu$m.

The plots of the field-emission current calculated with the use of the parameters listed here are given in Fig. 7. The calculations are done according to Eqs. (6), (14), and (15), and the Fowler-Nordheim formula for the partial current, $I_N = ASF_N^2 \exp(-\mathcal{F}/F_N)$. Since the calculated radial electric field in the region of the tip weakly depends on $z$ (in contrast to the case of the nanocluster studied above) and sharply decreases in the region of transition to the cylindrical part of the wire, the effective emitting area $S$ is estimated by the total area of the hemispherical tip, $S=2\pi R^2$. The work function is taken for silicon, $W=4.2$ eV. Next, the Fermi energy standing in the expression for $A$, see Eq. (2), is estimated from the equation $\varepsilon_F \simeq |e|F_{in}r_{TF}$, where $F_{in} \simeq \beta F_0/\epsilon$ is the field inside the semiconductor near the end of the tip, $r_{TF}$ is the Thomas-Fermi screening length, and $\epsilon$ is the dielectric constant of the semiconductor. Such an estimate, carried out for $n$-Si, leads to $\varepsilon_F \simeq 0.1$ eV at $F_0 \simeq 2\times 10^6$ V/cm. The picture of Coulomb staircase shown in Fig. 7 is basically the same as that in Fig. 4. Again, the increase of the current with the applied field is determined by the increase of the effective field (14) due to both single-electron charging (steps) and charge polarization under a constant charge (regions between the steps). The main difference is that the interval of the applied field needed for addition of one electron to the wire is considerably reduced, owing to the larger capacitance $C_2$, and appears to be of about 1.2 V/$\mu$m (further reduction of this interval takes place with the increase of the wire length, see below). Next, since the capacitance $C$ increases considerably in comparison to the case of nanocluster of the same radius, the Coulomb blockade features at room temperature are poorly visible even in the derivative plot, see the inset. Nevertheless, these features remain pronounced at $T=77$ K.

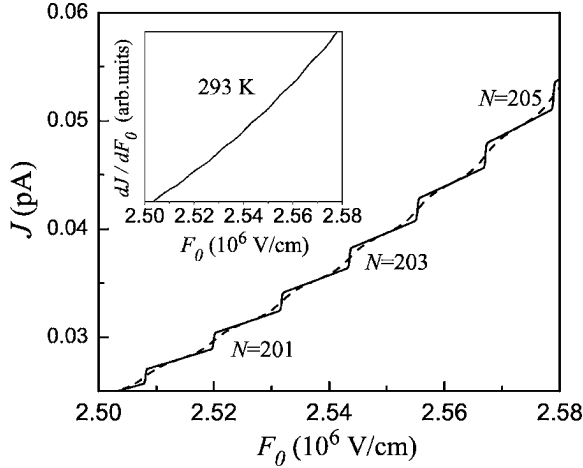

FIG. 7. Current from the cylindrical wire of radius $R=5$ nm and length 0.1 $\mu$m as a function of the applied field $F_0 = -V/L$ at $T=4.2$ K (solid) and 77 K (dashed). The inset shows the derivative of the current at $T=293$ K.

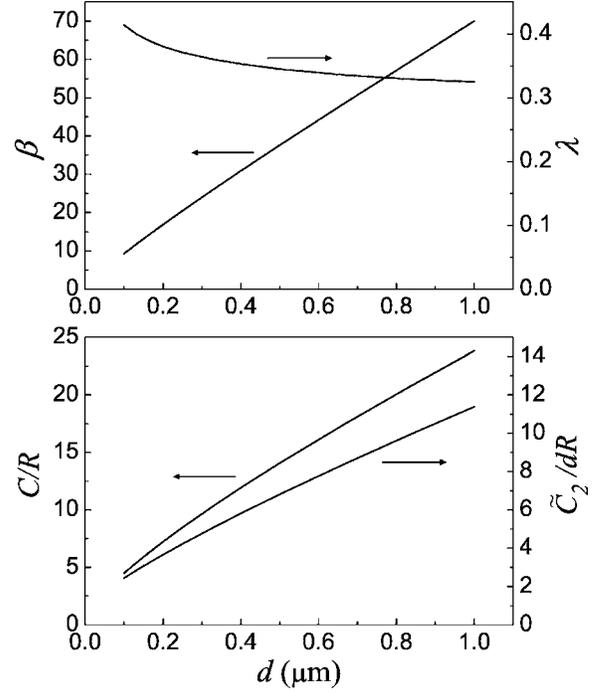

FIG. 8. Dependence of the parameters $\beta$, $\lambda$, $C$, and $\tilde{C}_2$ on the length of the wire for $R=5$ nm and $F_0=5\times 10^5$ V/cm.

With the increase of the wire length $d$, the parameters entering Eqs. (14) and (15) are modified as shown in Fig. 8. The field enhancement factor and the capacitances increase nearly in a linear way, while the parameter $\lambda$, which characterizes relative contribution of charging to the effective field, slightly decreases (for comparison, the idealized emitter considered in the previous section is described by the parameters $\beta=d/R$, $\lambda=1$, $C=R$, and $\tilde{C}_2=dR$, where $d$ is the distance from the cathode to the emitting sphere). The increase of the total capacitance $C$ makes it difficult to observe the Coulomb staircase in long wires. For example, at $d=1$ $\mu$m one should have temperatures considerably lower than 77 K. The interval of the applied field corresponding to the addition of one electron is inversely proportional to $\tilde{C}_2$. This interval decreases very fast with the increase of $d$ and becomes equal to $2.5\times 10^2$ V/cm at $d=1$ $\mu$m.

## IV. CONCLUSIONS

The key point of the presented theoretical study is the possibility of noticeable modification of the effective electric field causing the field emission from a nanoscale conductor by addition of just one electron to this conductor. Formally, this modification is described by introducing the effective field $F_N$, which determines the partial current $I_N$, and by evaluating the dependence of this field on the bias applied between the cathode and anode, see Eqs. (11) and (14). As a result of this effect, the current-voltage characteristics of the field emission show steps in the Coulomb blockade regime. In other words, the steplike current-voltage characteristics related to single-electron charging (Coulomb staircases) may exist even under the conditions of field-emission experiments, when the applied bias is orders of magnitude larger than the charging energy. The steps on the current-voltage characteristics can be visible at 77 K in the case of field emission from nanoclusters and nanowires of 10 nm diameter and submicron length. In the regions between the steps, where the total charge of the nanoscale object is constant, the current increases with the increase of the applied bias owing to charge polarization.

The staircases described in this work are similar to the usual Coulomb staircases obtained in the transport through small metallic islands[28–30] or quantum dots (see Ref. 35 for review) with strong asymmetry in the barriers. In both cases, each step of the current is associated with addition of an electron to the nanoscale object, and the applied source-drain voltage drops mostly across the low-transparent barrier (the barrier between the object and the drain). Therefore, the periodicity of the steps in both cases is determined by the object-drain capacitance $C_2$. However, the steps in the second case are formed due to shifts of effective ($N$-dependent) electrochemical potential of the object with respect to electrochemical potentials of the source and drain. For this reason, the usual Coulomb staircase shows well-defined steps when $C_2$ is greater than the object-source capacitance $C_1$, while in the opposite situation, $C_1 \gg C_2$, the steps are suppressed and the current-voltage characteristic approaches to a linear dependence.[29,30] In contrast, in the case described in this work the steps are formed due to changes in the probability of Fowler-Nordheim tunneling from the object to the drain (anode). That is why the steps are clearly visible under the condition $C_1 \gg C_2$, which is imposed by the field-emission layout considered in this paper. To summarize, the sensitivity of the field emission to the number of electrons in the nanoscale object makes it possible to obtain the Coulomb staircases under the conditions when such staircases cannot be observed in the transport through small metallic islands or quantum dots.

The quantitative consideration has been applied here to some simple models of the nanoscale objects, whose electrostatic properties necessary for description of the field enhancement and charging have been determined consistently. Consequently, the number of geometrical parameters characterizing the objects has been minimized. For example, the nanowire has been characterized only by its length $d$ and radius $R$. In reality, the geometrical structure of objects is more complicated. For example, their tips may contain sharp regions which provide a more efficient field emission. In fact, high field-emission currents from nanoscale objects are typically observed at the applied fields of the order of $10^5$ V/cm, which requires the field enhancement factors much larger than those calculated in this paper. On the other hand, the presence of sharp tips cannot strongly modify the capacitances of the objects. The general picture of the single-electron tunneling under the field-emission regime also remains valid. For possible application to experiments, the field enhancement due to charging can be described by equations of the kind of Eqs. (11) and (14), where $\beta$ and $\lambda$ should be considered as parameters to be determined experimentally.

At the present time, there is no experimental evidence of the Coulomb staircase phenomenon under the field emission. Though the current-voltage characteristics sometimes show steplike features, see, for example, Ref. 11, these features are not regular and, most probably, should be attributed to instabilities of the emission process and burning out of the emitting material. There are numerous reasons which make observation of the phenomena considered in this paper difficult. First of all, in most cases the nanoscale objects on the cathode surface form dense arrays. This means that the field emission takes place from a macroscopic number of objects which are electrostatically coupled. The charging and field-emission properties appear to be considerably different[36] from those of individual objects. The Coulomb blockade phenomena in this case should be dramatically suppressed by the size dispersion of the objects and by the effects of mutual screening. Investigation of field emission from individual objects is possible in the cases of metallic nanoclusters[8–10] and carbon nanotubes.[26] However, there exists the problem of electric isolation of these objects from the cathode, which is one of the necessary conditions for Coulomb blockade. No special attempts to achieve such an isolation in the field-emission experiments have been undertaken so far, except for the nanomechanical system investigated in Ref. 7, where the electron emission from an isolated Au island to a submicron-sized electrode has been observed. Most of the experiments on field emission are carried out at room temperature, though existing experimental techniques also allow measurements at liquid nitrogen temperature. This means that the Coulomb blockade phenomena can be observed only for small-sized objects whose capacitances are low enough (see the results of Sec. III). Besides, the interval of the applied field corresponding to addition of one electron strongly decreases in the case of emission from long nanowires, which requires high resolution with respect to field. In summary, a search for the Coulomb blockade features in the field-emission current would require a special planning of experiment. The author hopes that the presented theoretical study will stimulate experimental investigations in this direction.

## ACKNOWLEDGMENTS

The author is grateful to A. I. Klimovskaya for stimulating discussions.

## APPENDIX: KERNEL OF EQUATION (13)

If $z<d$ and $z'<d$, $K(z,z')=K_0(z,z')-K_0(-z,z')$, where

$$K_0(z,z') = \int_0^\pi \frac{d\varphi}{\pi} \frac{|e|}{\sqrt{(z-z')^2 + 2R^2(1-\cos\varphi)}}. \tag{A1}$$

If $z<d$ and $z'>d$, $K(z,z')=K_0(z,z')-K_0(-z,z')$, where

$$K_0(z,z') = \int_0^\pi \frac{d\varphi}{\pi} \frac{|e|}{\sqrt{(d-z)^2 + 2R(d-z)\cos\theta' + 2R^2(1-\sin\theta'\cos\varphi)}}. \tag{A2}$$

If $z>d$ and $z'<d$, $K(z,z')=K_0(z,z')-K_0(z,-z')$, where

$$K_0(z,z') = \int_0^\pi \frac{d\varphi}{\pi} \frac{|e|}{\sqrt{(d-z')^2 + 2R(d-z')\cos\theta + 2R^2(1-\sin\theta\cos\varphi)}}. \tag{A3}$$

Finally, if $z>d$ and $z'>d$,

$$K(z,z') = \int_0^\pi \frac{d\varphi}{\pi} \left[ \frac{|e|}{\sqrt{2R^2(1-\cos\theta\cos\theta' - \sin\theta\sin\theta'\cos\varphi)}} \right.$$
$$\left. - \frac{|e|}{\sqrt{4d^2 + 4dR(\cos\theta+\cos\theta') + 2R^2(1+\cos\theta\cos\theta' - \sin\theta\sin\theta'\cos\varphi)}} \right]. \tag{A4}$$

In Eqs. (A2)–(A4), $\cos\theta=(z-d)/R$ and $\cos\theta'=(z'-d)/R$, so $\theta$ and $\theta'$ are the azimuthal angles. The integrals are taken over the polar angle $\varphi$. The function $K(z,z')$ is also representable in the form of full elliptic integrals.

---